\documentclass[a4paper]{jpconf}
\usepackage{graphicx}

\usepackage{hyperref}

\begin{document}
\title{Dark matter Axion search with riNg Cavity Experiment DANCE: Design and development of auxiliary cavity for simultaneous resonance of linear polarizations }

\author{Hiroki Fujimoto$^1$, Yuka Oshima$^1$, Masaki Ando$^{1,2}$,Tomohiro Fujita$^{2,3}$,\\
Yuta Michimura$^{1,4}$, Koji Nagano$^5$ and Ippei Obata$^6$}

\address{$^1$Department of Physics, University of Tokyo, Bunkyo, Tokyo 113-0033, Japan\\
$^2$Research Center for the Early Universe, University of Tokyo, Bunkyo, Tokyo 113-0033, Japan\\
$^3$Waseda Institute for Advanced Study, Waseda University, Shinjuku, Tokyo 169-0051, Japan\\
$^4$PRESTO, Japan Science and Technology Agency (JST), Kawaguchi, Saitama 332-0012, Japan\\
$^5$Institute of Space and Astronautical Science, Japan Aerospace Exploration Agency, Sagamihara, Kanagawa 252-5210, Japan\\
$^6$Max-Planck-Institut für Astrophysik, Karl-Schwarzschild-Straße 1, 85741 Garching, Germany}

\ead{hiroki.fujimoto@phys.s.u-tokyo.ac.jp}

\begin{abstract}
Axion-like particles (ALPs) are undiscovered pseudo-scalar particles that are candidates for ultralight dark matter. 
ALPs interact with photons slightly and cause the rotational oscillation of linearly polarized light. 
Dark matter Axion search with riNg Cavity Experiment (DANCE) searches for ALP dark matter by amplifying the rotational oscillation with a bow-tie ring cavity. 
Simultaneous resonance of linear polarizations is necessary to amplify both the carrier field and the ALP signal, and to achieve the design sensitivity. 
The sensitivity of the current prototype experiment DANCE Act-1 is less than expectation by around three orders of magnitude due to the resonant frequency difference between s- and p-polarization in the bow-tie ring cavity.
In order to tune the resonant frequency difference, the method of introducing an auxiliary cavity was proposed. 
We designed an auxiliary cavity that can cancel out the resonant frequency difference and realize simultaneous resonance, considering optical loss.
We also confirmed that the sensitivity of DANCE Act-1 with the auxiliary cavity can reach the original sensitivity.
\end{abstract}

\section{Introduction}
Axion-like particles (ALPs) are undiscovered pseudo-scalar particles predicted from high energy physics such as string theory.
ALPs have a broad range of mass and can be treated as coherent classical fields due to their small mass, $m_a\ll$ eV.
ALPs produced via the misalignment mechanism behave as dark matter by oscillating its background field \cite{JP, LF, MD,PA} and are considered to be candidates for dark matter. The oscillating ALP field causes a small phase velocity difference between left- and right-handed circularly polarized light \cite{SM1, SM2}, given by

\begin{equation}
	c_{\rm{L/R}}(t) = \sqrt{1 \pm \frac{g_{a\gamma} a_0 m_a}{k} \sin (m_a t + \delta_{\tau}(t))},
\end{equation}
where $g_{a\gamma}$ is the axion-photon coupling constant, $a_0$ is the constant amplitude, $k$ is the wave number of the photon, and $\delta_{\tau}(t)$ is the phase factor.
From the point of view of linear polarizations, this phase velocity difference can be regarded as the rotational oscillation of the polarization plane, and the signal of ALP dark matter appears as the sidebands with the polarization orthogonal to the carrier polarization \cite{HL}.

When the linear polarization travels through the oscillating ALP field, the sidebands produced by ALP are amplified until the optical path reaches the length of the half period of the ALP oscillation.
Therefore,  the ALP signal can be amplified by extending the light path using optical cavities.
However, in a linear cavity, the rotation is cancelled out during one round trip due to the polarization flip at the reflection in the mirrors.
To deal with this effect, we proposed Dark matter Axion search with riNg Cavity Experiment (DANCE) \cite{DANCE, DANCE2}. 
DANCE makes use of an optical bow-tie ring cavity with four mirrors.
Two reflections at each end of the cavity prevent the polarization flip and can amplify the rotational oscillation during one round trip.

Currently, the prototype experiment: DANCE Act-1 is underway for identifying technical issues and demonstrating the proof of its principle.
So far, the resonant frequency difference of $\sim$3 MHz between s-polarization (s-pol.) and p-polarization (p-pol.) has been observed in the cavity of DANCE Act-1 \cite{DANCE_oshima_moriond}.
This resonant frequency difference degrades the sensitivity of DANCE Act-1.
One of the ways to realize simultaneous resonance is to install an auxiliary cavity for cancelling out the reflective phase difference \cite{Auxiliary}.
However, installing an auxiliary cavity can produce additional loss due to the mode mismatch and anti-reflective (AR) coatings of the mirrors.
Optical loss in the main cavity leads to the degradation of the sensitivity as well.
In this paper, we report on the design of the auxiliary cavity for the DANCE Act-1, considering the feasibility of simultaneous resonance and the additional loss from the auxiliary cavity.

\section{Resonant frequency difference in DANCE Act-1}
In DANCE Act-1, the resonant frequency difference of $\sim$3 MHz between s- and p-pol. is observed.
This resonant frequency difference is produced by the oblique incidence on the cavity mirrors.
When there is resonant frequency difference in the bow-tie ring cavity, only either the carrier polarization or the sidebands polarization can be resonant and amplified in the cavity.
Due to this non-simultaneous resonance, the target sensitivity of DANCE Act-1 is degraded by three orders of magnitude (Figure \ref{sensitivity}).
To improve the sensitivity, the resonant frequency difference needs to be cancelled out without introducing excessive optical loss in the cavity.

\section{Designing the auxiliary cavity for simultaneous resonance}

\begin{figure}[t]
\begin{minipage}{16pc}
\includegraphics[width=14pc]{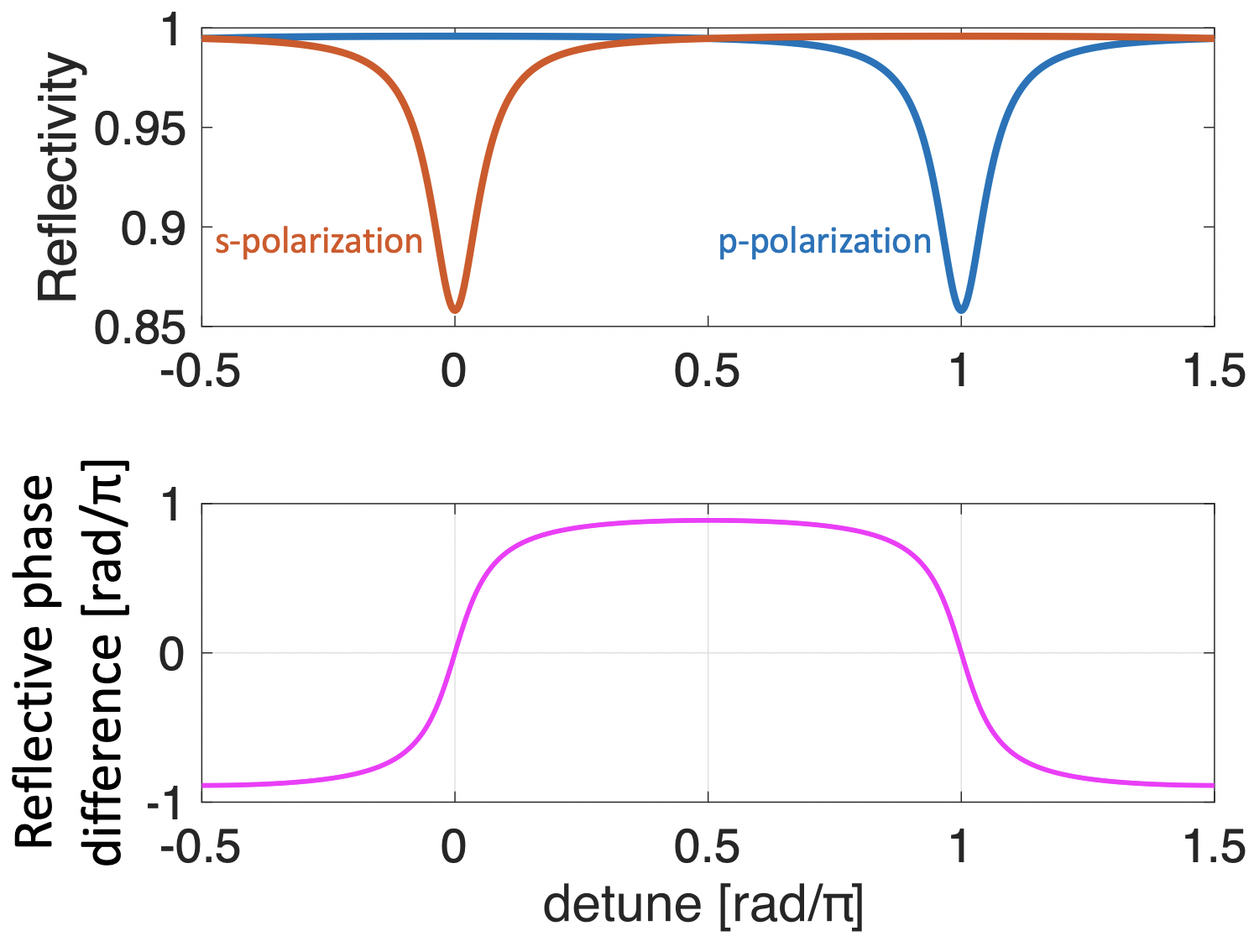}
\caption{\label{label}An example of reflection by a detuned auxiliary cavity. The lower graph shows the reflective phase difference between s- and p-pol.} \label{aux_ref}
\end{minipage}\hspace{2pc}%
\begin{minipage}{20pc}
\includegraphics[width=20pc]{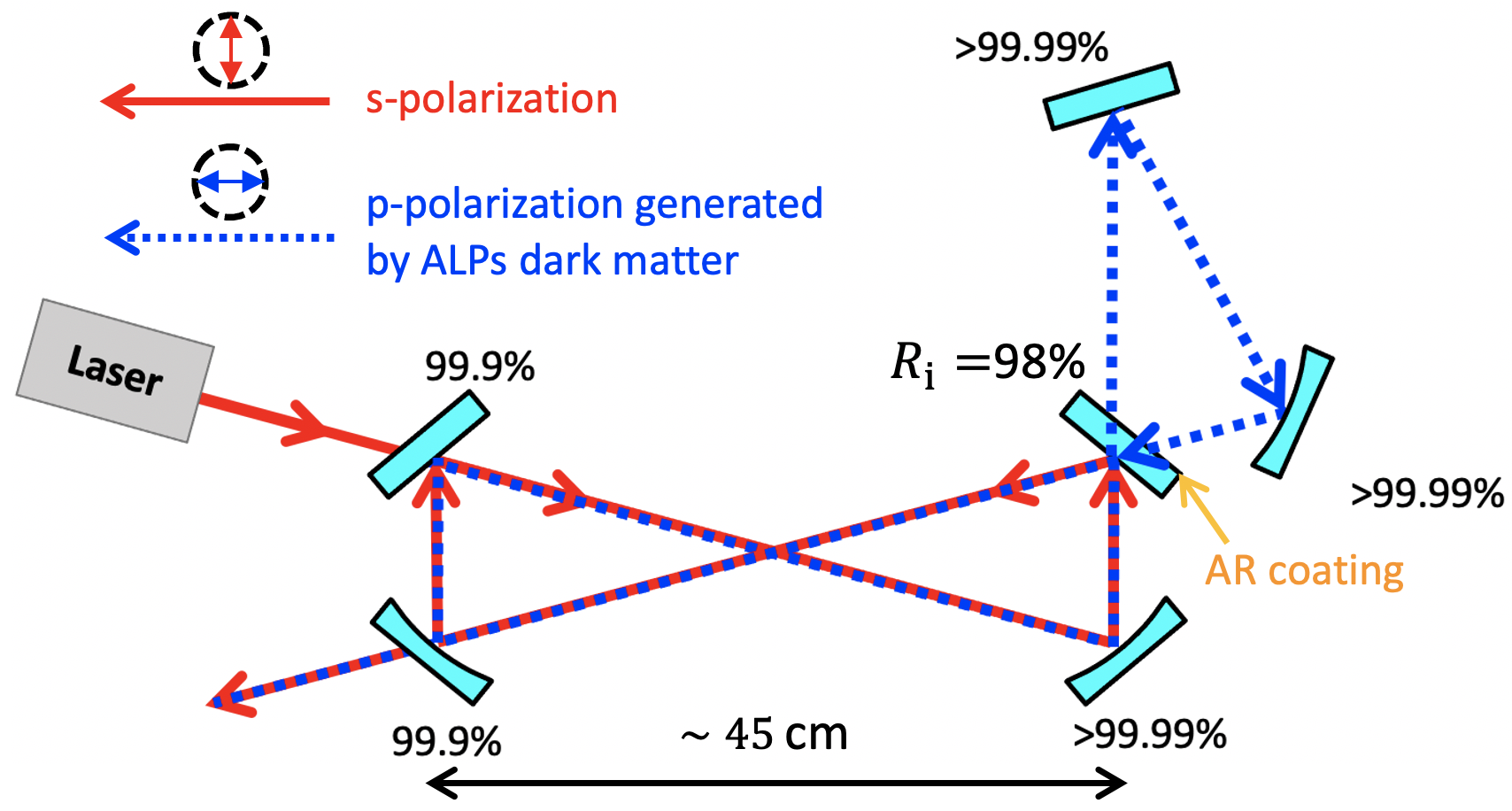}
\caption{\label{label}Configuration of DANCE Act-1 with the auxiliary cavity. The triangle auxiliary cavity is attached to the main bow-tie ring cavity.} \label{configuration}
\end{minipage} 
\end{figure}

To realize simultaneous resonance, the method of attaching an auxiliary cavity to the main cavity has been proposed \cite{Auxiliary}.
By constructing an auxiliary cavity with an odd number of mirrors, the resonance of s- and p-pol. can be separated by $\pi$ rad in the auxiliary cavity.
In addition, the reflective phase shift of a cavity changes drastically at the resonant frequency.
With these characteristics, the reflective phase difference between s- and p-pol. of the auxiliary cavity can be controlled by tuning the resonant condition of the cavity (Figure \ref{aux_ref}).
Therefore, by attaching this auxiliary cavity to the main cavity for cancelling out the resonant frequency difference with the reflective phase difference of the auxiliary cavity, simultaneous resonance can be realized.

\begin{figure}[t]
\includegraphics[width=38pc]{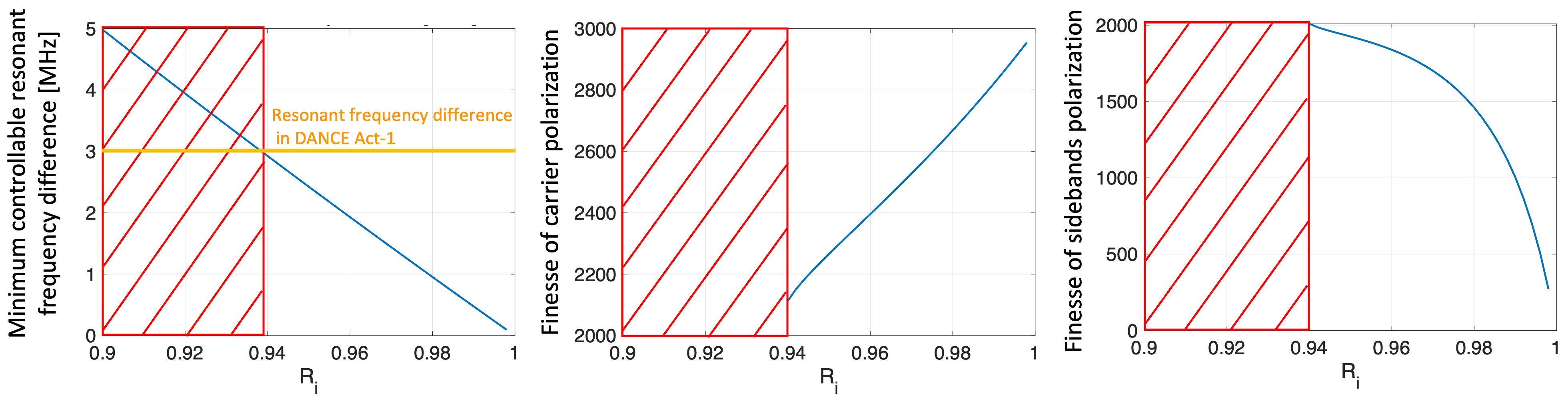}\hspace{2pc}%
\caption{\label{label}Various parameters of DANCE Act-1 with varying power reflectivity of the intermediate mirror $R_{\rm{i}}$ with an auxiliary cavity. Achieving simultaneous resonance is unfeasible in the red shaded area due to the minimum controllable resonant frequency difference being larger than the measured value in DANCE Act-1.} \label{feasibility}
\end{figure}

In order to introduce the technique of an auxiliary cavity to DANCE Act-1, the configurations of the main bow-tie ring cavity and the triangle auxiliary cavity were designed (Figure \ref{configuration}).
The plane mirrors and the concave mirrors are arranged so that the transverse eigenmode of the main cavity matches that of the auxiliary cavity.
The power reflectivity of the intermediate mirror $R_{\rm{i}}$ between the main cavity and the auxiliary cavity influence the feasibility of simultaneous resonance and the sensitivity to ALP dark matter.
When higher $R_{\rm{i}}$ is designed, the range of the reflective phase difference of the auxiliary cavity becomes wider and smaller resonant frequency difference can be cancelled out.
However, at the same time, it increases the finesse, which corresponds to the number of round trips in the cavity, and enhances the optical loss of the AR coating.
This effect degrades the finesse of the main cavity and the sensitivity to ALP dark matter.
On the other hand, by decreasing $R_{\rm{i}}$, the degradation of the finesse of the main cavity can be suppressed at the expense of the feasibility of achieving simultaneous resonance.

The value of $R_{\rm{i}}$ was determined so that the minimum controllable resonant frequency is smaller than the resonant frequency difference of DANCE Act-1.
In order to estimate the feasibility of achieving simultaneous resonance and finesse of the main cavity, the complex reflectivity of the auxiliary cavity for s- and p-pol were calculated as 

\begin{equation}
  r_{\rm{aux. s/p}} = \pm \sqrt{R_{\rm{i}}} + \frac{M (1-R_{\rm{i}}) \sqrt{1-T_{\rm{loss}}} e^{-i\phi}}{1 \pm  \sqrt{R_{\rm{i}}} \sqrt{1-T_{\rm{loss}}} e^{-i\phi}},
\end{equation}
where $M$ is the mode-matching ratio between the main and the auxiliary cavity,  $T_{\rm{loss}}$ is the loss of the AR coating, and $\phi$ is the round-trip phase shift of the auxiliary cavity.

Figure \ref{feasibility} shows the minimum controllable resonant frequency difference and the finesse for s- and p-pol. in the main cavity with different $R_{\rm{i}}$.
Here we assumed $M = 0.99$, $T_{\rm{loss}} = 0.01$ and the resonant frequency difference of 3 MHz.
To provide a margin to the minimum controllable resonant frequency difference, $R_{\rm{i}}$ was finally chosen to be 0.98.
With this reflectivity, the finesse of $\sim$ 2700 and $\sim$ 1500 for s- and p-pol. can be achieved, respectively.

\begin{figure}[t]
\includegraphics[width=18.5pc]{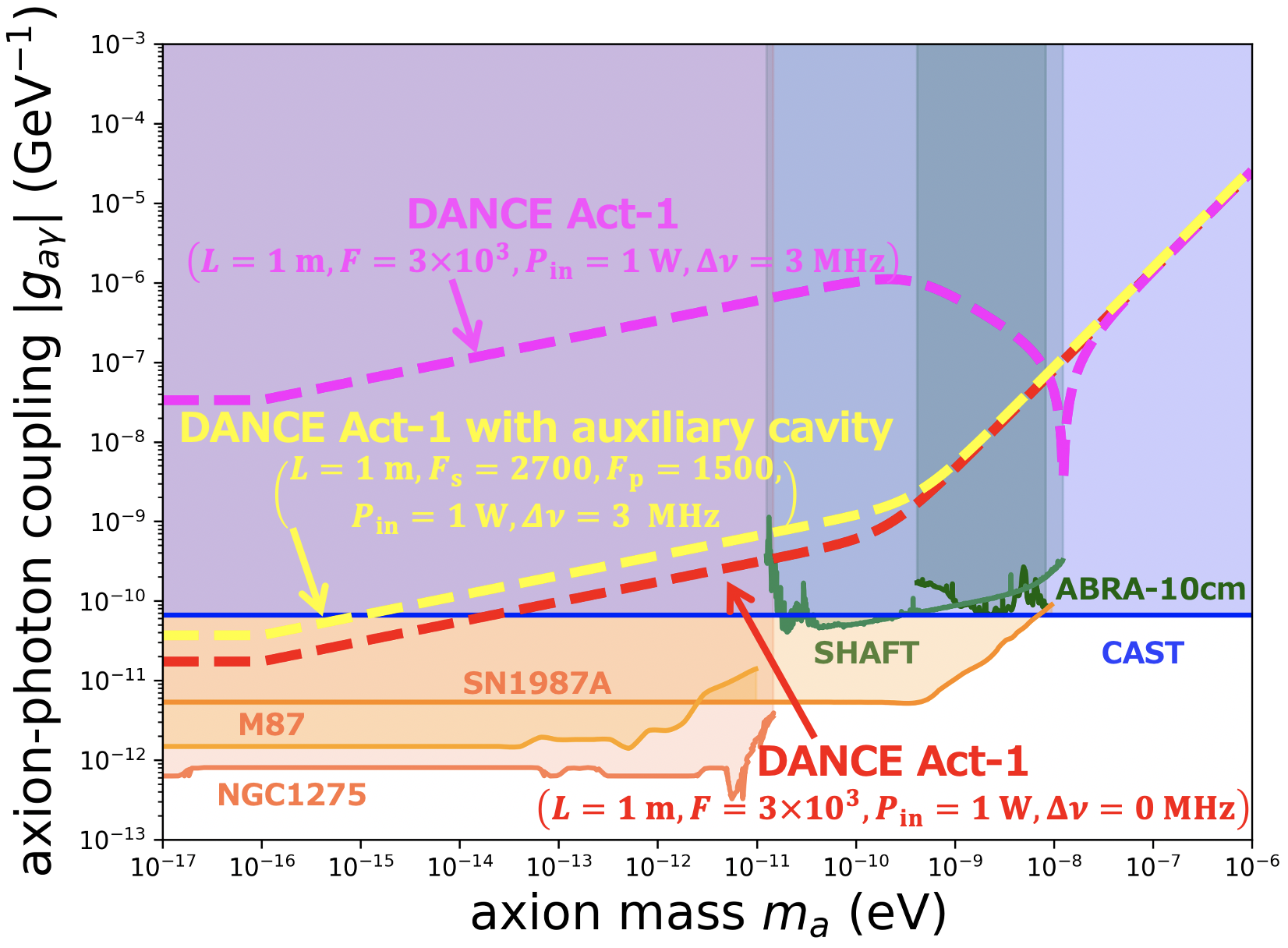}\hspace{1pc}%
\begin{minipage}[b]{18.5pc}\caption{\label{label}The sensitivity curves and the constraints on the axion-photon coupling $g_{a\gamma}$. The shot noise limits of DANCE Act-1 with the integration time of a year are shown as dotted lines.
The solid line shows the current bounds that are obtained from the experimental results of CAST \cite{CAST}, SHAFT \cite{SHAFT} and ABRACADABRA-10cm \cite{ABRA-10cm}, and from the astrophysical constraints by the gamma-ray observations of SN1987A \cite{SN1987A} and the X-ray observations of the M87 galaxy \cite{M87} and the NGC1275 galaxy \cite{NGC1275}.} \label{sensitivity}
\end{minipage}
\end{figure}

Figure \ref{sensitivity} shows the shot noise limited sensitivity of DANCE Act-1 with the designed auxiliary cavity.
By introducing the auxiliary cavity, the current target sensitivity is improved by around three orders of magnitude and reaches the initial target sensitivity.
At the time of writing, the assembly of the auxiliary cavity and the development of the control scheme are underway.

\section{Conclusion}
DANCE aims to search for ALP dark matter with an optical bow-tie ring cavity.
The current prototype experiment DANCE Act-1 has the resonant frequency difference of $\sim$3 MHz between s-and p-pol. and loses around three orders of magnitude of sensitivity.
To realize simultaneous resonance, we designed an auxiliary cavity considering  the optical loss and the feasibility of simultaneous resonance.
With the designed configuration, the sensitivity can reach the initial target sensitivity with no resonant frequency difference.
The assembly of the auxiliary cavity and the development of the control scheme are underway.

\ack
We would like to thank Shigemi Otsuka and Togo Shimozawa for manufacturing the mechanical parts of the bow-tie ring cavity. We also thank Ooi Ching Pin for editing this document. This work was supported by JSPS KAKENHI Grant Nos. 18H01224, 20H05850, 20H05854 and 20H05859, and JST PRESTO Grant No. JPMJPR200B.

\end{document}